\documentclass[prd,aps,twocolumn,a4paper,floatfix]{revtex4}

\usepackage{graphicx,psfrag}
\usepackage{amsmath,amsfonts,amssymb,amsthm}
\usepackage[mathscr]{euscript}
\usepackage{hyperref}

\usepackage{float}
\usepackage{upgreek}
\usepackage{bbm}

\newcommand{\non}{\nonumber}

\newcommand{\p}{\partial}

\newcommand{\vhat}{\hat{v}}

\newcommand{\cs}{c_{s}}

\newcommand{\pb}{\bot{b}}

\newcommand{\gamN}{{}^{\textrm{\tiny{(N)}}}\!\gamma}
\newcommand{\gamu}{{}^{\textrm{\tiny{(u)}}}\!\gamma}
\newcommand{\gB}{\mathbbmss{g}}
\newcommand{\gBi}{(\mathbbmss{g}^{-1})}
\newcommand{\gBN}{{}^{\textrm{\tiny{(N)}}}\!\mathbbmss{g}}
\newcommand{\gBNi}{{}^{\textrm{\tiny{(N)}}}\!(\mathbbmss{g}^{-1})}

\newcommand{\s}{{\mathbbmss{s}}}

\newcommand{\shat}{{\hat{\mathbbmss{s}}}}

\newcommand{\Qa}{{{Q}_1}}
\newcommand{\Qb}{{{Q}_2}}

\newcommand{\orta}{{A}}
\newcommand{\ortb}{{B}}
\newcommand{\ortc}{{C}}

\newcommand{\ortA}{{\mathbbmss{A}}}
\newcommand{\ortB}{{\mathbbmss{B}}}
\newcommand{\ortC}{{\mathbbmss{C}}}

\newcommand{\ortAhat}{{\hat{\mathbbmss{A}}}}
\newcommand{\ortBhat}{{\hat{\mathbbmss{B}}}}
\newcommand{\ortChat}{{\hat{\mathbbmss{C}}}}

\newcommand{\perpqn}{{}^{{\tiny{q}}}\!\!\!\perp}
\newcommand{\perpq}{{}^{{\tiny{\mathbbmss{q}}}}\!\!\!\perp}

\newcommand{\perpQ}{{}^{\textrm{{\tiny{Q}}}}\!\!\!\perp}

\newcommand{\epsu}{{}^{\textrm{{\tiny{(u)}}}}\!\epsilon}
\newcommand{\epsuS}{{}^{\textrm{{\tiny{(S)}}}}\!\epsilon}

\begin{document}

\title{Hyperbolicity of Divergence Cleaning and Vector Potential
  Formulations of GRMHD}

\author{David \surname{Hilditch}$^{1}$ and Andreas 
\surname{Schoepe}$^{2}$}

\affiliation{${}^1$CENTRA, Departamento de F\'isica, Instituto Superior
T\'ecnico – IST, Universidade de Lisboa – UL, Avenida Rovisco Pais 1, 1049
Lisboa, Portugal,\\
${}^2$Friedrich-Schiller-Universit\"at Jena, 07743 Jena, Germany.}

\date{\today}

\begin{abstract}
We examine hyperbolicity of general relativistic magnetohydrodynamics
with divergence cleaning, a flux-balance law form of the model not
covered by our earlier analysis. The calculations rely again on a
dual-frame approach, which allows us to effectively exploit the
structure present in the principal part. We find, in contrast to the
standard flux-balance law form of the equations, that this formulation
is strongly hyperbolic, and thus admits a well-posed initial value
problem. Formulations involving the vector potential as an evolved
quantity are then considered. Carefully reducing to first-order, we
find that such formulations can also be made strongly hyperbolic.
Despite the unwieldy form of the characteristic variables we therefore
conclude that of the free-evolution formulations of general
relativistic magnetohydrodynamics presently used in numerical
relativity, the divergence cleaning and vector potential formulations
are preferred.
\end{abstract}

\maketitle

\section{Introduction}\label{section:Introduction}

It is well appreciated~\cite{BauSha10x,Shi16} that the numerical
modeling of binary neutron star spacetimes plays, and will continue to
play, an important role in the new field of gravitational wave
astronomy, particularly in the case of multimessenger events. Such
simulations are, however, hampered by relatively poor error behavior
as compared with their vacuum, black hole counterparts. This is in
part because the equations of motion of these models have a more
complicated structure than those of pure general relativity, and are
hence less well understood, but also because solutions naturally
develop nonsmooth features, not to mention the ever-present
complication of the stellar surface.

In a recent contribution~\cite{SchHilBug17} we employed a new tool,
the dual-frame (DF) formalism~\cite{HilRic13,Hil15,HilRui16,HilHarBug16},
to analyze well-posedness of various fluid models. Well-posedness is
the weakest necessary condition to require of a set of evolution
partial differential equations (PDE) so that numerical approximation
to their solutions may be meaningfully sought. The formalism can be
used to exploit structure in the field equations and hence simplifies
earlier treatments. This should allow more sophisticated results to be
shown in the future.

One of the models treated in Ref.~\cite{SchHilBug17} was (ideal)
general relativistic magnetohydrodynamics (GRMHD), taken in two
different guises. In the Valencia flux-balance law
form~\cite{AntZanMir05} we found that the field equations are only
weakly hyperbolic, and therefore have an ill-posed initial value
problem. Here we attend to two flavors of GRMHD untouched by our
earlier study, namely the hyperbolic divergence cleaning (HDC) and
vector potential (VP) formulations. Our main result is that both are
strongly hyperbolic, provided suitable choices are made in the
first-order reduction of the latter.

We work in~$3+1$ dimensions in geometric units with~$c=G=1$. Our
calculations were performed primarily with xTensor for {\it
  Mathematica}~\cite{xAct_web_aastex}; our notebooks are available
online in Ref.~\cite{HilSch18_WebTar}.

\section{Mathematical background}\label{section:MathBackground}

We start with a short overview of the relevant theory, definitions,
and results to the PDE analysis and the DF
formalism. These are taken in a highly summarized form
from Refs.~\cite{Hil15,HilHarBug16,SchHilBug17}.

\paragraph*{Index notation.} Latin letters~$a$--$e$ are used as abstract
indices. We also use~$p$ as an abstract index, placing it always on
the spatial derivative appearing on the right-hand side of our
first-order PDE system.  The four-metric~$g_{ab}$ is the only object
permitted to raise and lower indices. The symbol~$\p_a$ stands for a
flat covariant derivative. Indices~$u$, $S$,~$s$,~$\shat$ and~$\s$ label
contraction in that slot with~$u^a$ or~$u_a$ and so on,
respectively. Capital Latin letters~$A$--$C$ are taken as abstract
indices and denote appliance of the projection operators~$\perpQ$ 
and~$\perpqn$, to
be defined later.  Similarly, we use indices~$\ortA$--$\ortC$
and~$\ortAhat$--$\ortChat$ to denote the application of the projection
operator~$\perpq$ over a vector or dual vector, respectively.

\paragraph*{DF formalism.} We describe a region of spacetime in two
different frames, namely the lowercase and the uppercase frame. We
take the lowercase frame as an Eulerian frame, associated with a
coordinate basis as is standard in numerical relativity. We denote the
future pointing timelike unit normal vector to spatial slices of
constant time, as usual, by~$n^a$. Additionally, we take any three
linearly independent vector fields orthogonal to~$n^a$ to form a basis
of the four-dimensional spacetime.  Tensors orthogonal to~$n^a$ are
called lowercase spatial, or just lowercase. The uppercase frame
consists of a future pointing timelike unit vector~$N^a$, which is
identified in the application below with the fluid
four velocity~$u^a$, plus any three linearly independent vector fields
orthogonal to~$N^a$. Tensors orthogonal to~$N^a$ are likewise called
uppercase spatial, or just uppercase.  The future pointing unit
vectors of the lower- and uppercase frames can be mutually~$3+1$
decomposed as
\begin{align}
n^a=W(N^a+V^a)\,,\qquad N^a=W(n^a+v^a)\,,
\end{align}
with the Lorentz factor~$W=(1-V^aV_a)^{-1/2}=(1-v^av_a)^{-1/2}$. The
vectors~$v^a=\hat{v}^a/W$ and~$V^a$ are the boost vectors orthogonal
to~$n^a$ and~$N^a$, respectively. We define projection operators by
\begin{align} 
\gamma^b{}_a=\delta^b{}_a+n^bn_a\,,\qquad
\gamN^b{}_a=\delta^b{}_a+N^bN_a\,,
\end{align}
which are also denoted as the lowercase and uppercase spatial metrics,
respectively. By definition, the relations~$\gamma^b{}_an_b=0$,
$\gamN^b{}_aN_b=0$ hold.  We define furthermore the lowercase and
uppercase boost metrics and their inverses, which are presented
in~Table~\ref{tab:threeplusonedecomp}.

\begin{table}[t]
\centering
\begin{tabular}{l|c|c}
\hline\hline
  & Uppercase & Lowercase\\
\hline
Unit normal  & $N^{a}=W(n^a+v^a)$ & $n^a=W(N^a+V^a)$ \\
Boost vector & $V^a$ & $\qquad v^a=\hat{v}^a/W$ \\ 
Lorentz factor& $W=(1-V^aV_a)^{-1/2}$ & $W=(1-v^av_a)^{-1/2}$ \\ 
Projector & $\gamN^a{}_b=g^a{}_b+N^aN_b$ & $\gamma^a{}_b=
g^a{}_b+n^an_b$\\ 
Boost metric & $\gBN_{ab} :=\gamN_{ab}+W^2 V_aV_b$ & 
$\gB_{ab} :=\gamma_{ab}+\vhat_a\vhat_b$\\ 
Inverse boost  & $\gBNi^{ab} =\gamN^{ab}-V^aV^b$ & 
$\gBi^{ab} =\gamma^{ab}-v^av^b$\\
\hline
\end{tabular}
\caption{Overview of the uppercase and lowercase
quantities.~\label{tab:threeplusonedecomp}}
\end{table}

\paragraph*{PDE analysis.} We consider a quasilinear system of
first-order evolution PDEs, in this case GRMHD with HDC, 
written in the form
\begin{align}
 \nabla_u \mathbf{U}=\mathbf{A}^p\nabla_p\mathbf{U}
 +\boldsymbol{\mathcal{S}}\,,\label{equation:PDEsystem}
\end{align}
with the covariant derivative along the streamlines of the fluid
elements~$\nabla_u\equiv u^a\nabla_a$ of the vector of evolved
variables, called the state vector~$\mathbf{U}$, on the left-hand
side. On the right-hand side, the covariant derivative of the state
vector is contracted with the principal
part~$\mathbf{A}^p$,~$\mathbf{A}^a u_a=0$. The
symbol~$\boldsymbol{\mathcal{S}}$ stands for the source term which
does not affect the level of hyperbolicity. We need only analyze the
system of evolution equations for the matter variables, since they are
minimally coupled to the Einstein equations for the components of the
metric tensor.

\paragraph*{Strong hyperbolicity.} For the hyperbolicity analysis,
we have to perform a~$2+1$ decomposition against lowercase and/or
uppercase spatial vectors and their respective orthogonal spatial
projectors. The relevant quantities are defined in
Table~\ref{tab:spatialvectors}. Taking an arbitrary uppercase unit
spatial 1-form~$S_a$, we define the uppercase principal symbol of the
system~\eqref{equation:PDEsystem} as
\begin{align}
\mathbf{P}^S\equiv\mathbf{A}^pS_p\,.
\end{align}
We call the system~\eqref{equation:PDEsystem} {\it weakly hyperbolic},
if for each~$S_a$ the eigenvalues of~$\mathbf{P}^S$ are real.  We call
the system~\eqref{equation:PDEsystem} {\it strongly hyperbolic}, if
the system is weakly hyperbolic and for each~$S_a$ the principal
symbol~$\mathbf{P}^S$ has a complete set of right eigenvectors written
as columns in a matrix~$\mathbf{T}_S$ and there exists a
constant~$K>0$, independent of $S_a$, such that~$\mathbf |\mathbf
T_S|+|\mathbf T_S^{-1}| \leq K$. Similar definitions are made if
  we~$3+1$ decompose the system against~$n^a$ rather than~$u^a$, and
  the initial value problem, where data are given at~$t=0$, can
  be well-posed only if it satisfies these {\it lowercase}
  strong hyperbolicity conditions~\cite{GusKreOli95,SarTig12,Hil13}.

\paragraph*{Frame and variable independence of hyperbolicity.} If the
uppercase eigenvalues of the principal symbol fulfill the
inequality~$|\lambda_{\text{N}}||V|<1$ then strong hyperbolicity is
independent of the chosen frame~\cite{SchHilBug17}. By the form of the
energy-momentum tensor of GRMHD, see below, a naturally preferred
frame is the fluid rest frame. Therefore, in the PDE analysis in
Sec.~\ref{section:HypGRMHDDivCl}, we work exclusively in the
uppercase frame, taken to be the fluid rest frame,~$N^a\equiv u^a$;
hence the~$3+1$ decomposition in Eq.~\eqref{equation:PDEsystem}, and in
the following, of the equations against the fluid four velocity~$u^a$
and the orthogonal projector~$\gamu^a{}_b$. In numerical applications,
particular sets of variables, such as the primitive or conservative
sets are used. In our analysis, we make a choice of variables which
differs slightly from those. Our variables are however related to the
code variables by a regular transformation, across which hyperbolicity
is unaffected.

\begin{table}[t]
\centering
\begin{tabular}{l|cc}
\hline\hline
& Uppercase & Lowercase  \\
\hline
Unit normal & $N^a$ & $n^a$   \\
Spatial 1-form & $S_a$ & $\s_a$ \\ 
Spatial vector &$S^a=\gamN^{ab}S_b$ &$\shat^a=\gBi^{ab}\s_{b}$ \\ 
Norm &$S_aS^a=1$ &$\s_a\gBi^{ab}\s_{b}=1$  \\
Projector& $\perpQ_{\ a}^{ b}=\gamN^{ b}{}_{a}-S^bS_a$ &
$\perpq_{\ a}^{ b}=\gamma_{\ a}^{ b}-\shat^b\s_a$ \\
Index notation & $\perpQ^\ortb \!\!_{ \orta}$  & $\perpq^\ortB \!\!_{ \ortAhat}$ \\ 
\hline
\end{tabular}
\caption{Summary of the various unit spatial vectors appearing in
our~$2+1$ decomposed equations, plus their associated projection
operators.~\label{tab:spatialvectors}}
\end{table}

\section{Basics of GRMHD}\label{section:BasicsGRMHD}

A brief review of the basic definitions, equations, and assumptions of
GRMHD with HDC is now given, following
Refs.~\cite{Ani90a,AntMirMar10,MoeMunFab14}. Presently, the focus
lies on the mathematical structure of the equations, and thus we
suppress some (important) physical insight and statements. We use
Lorentz-Heaviside units for electromagnetic quantities
with~$\varepsilon_0=\mu_0=1$, where~$\varepsilon_0$ and~$\mu_0$ are
the vacuum permittivity (or electric constant) and permeability (or
magnetic constant), respectively. Motivated by the arguments given in
the previous section, we work exclusively in the uppercase (fluid)
frame and thus, present the system of equations in a form so adjusted.

The energy-momentum tensor of GRMHD consists of an ideal fluid part,
\begin{align}
T^{ab}_{\text{fluid}}=\rho_0 h u^a u^b+g^{ab} p\,,
\label{equation:EnergyMomentumTensorHD}
\end{align}
with the four velocity of the fluid elements~$u^a$, rest mass
density~$\rho_0$, specific enthalpy~$h$, and pressure~$p$; plus the
standard electromagnetic energy-momentum tensor
\begin{align}
T^{ab}_{\text{em}}=F^{ac}F^{b}{}_{c}-
\frac{1}{4}g^{ab}F_{cd}F^{cd}\,,
\label{equation:EnergyMomentumTensorEM}
\end{align}
with the Faraday electromagnetic tensor field (or for short field
strength tensor)~$F^{ab}$.  The specific enthalpy~$h$ can be expressed
in terms of~$\rho_0,\ p,$ and the specific internal
energy~$\varepsilon$ as
\begin{align}
h = 1+\varepsilon +\frac{p}{\rho_0}\,.
\end{align}
The local speed of sound~$\cs$ is defined by the relation
\begin{align}
\cs^2=\frac{1}{h}\left( \chi+\frac{p}{\rho_0^2}\kappa
\right),\ \ \chi= \left( \frac{\p p}{\p \rho_0} \right)_{\varepsilon},\
\kappa=\left(\frac{\p p}{\p
\varepsilon}\right)_{\rho_0}\,.
\end{align}
We assume an equation of state (EOS) of the form
\begin{align}
p=p(\rho_0,\varepsilon) \label{equation:EOS},
\end{align}
with~$p>0$ given, satisfying furthermore that the local speed of
sound lies in the range~$0<\cs\leq1$.

Using the ideal MHD condition, where the electric conductivity tends
to infinity while the electric four-current remains bounded, the field
strength tensor and its dual become
\begin{align}
F^{ab}&=\epsilon^{abcd}u_c b_d\,,
\label{equation:FieldStrengthTensor}\\
{^*F^{ab}}&=u^a b^b-u^b b^a\,\label{equation:DualFieldStrengthTensor},
\end{align}
respectively, where we introduced the uppercase magnetic field
vector~$b^a$, satisfying $u_ab^a=0$, and the Levi-Civit\`a tensor
\begin{align}
\epsilon^{abcd}=-\frac{1}{\sqrt{-g}}\left[ a b c d \right]\,,
\end{align}
where~$g$ is the determinant of the spacetime metric~$g_{ab}$, $\left[
 a b c d \right]$ is the completely antisymmetric Levi-Civit\`a
symbol, and~$2 {^*F^{ab}}=-\epsilon^{abcd}F_{cd}$ holds. Note that we
use the sign convention of Ref.~\cite{AlcDegSal09}.

Taking the sum of Eqs.~\eqref{equation:EnergyMomentumTensorHD}
and~\eqref{equation:EnergyMomentumTensorEM}, and substituting the
field strength tensor~\eqref{equation:FieldStrengthTensor}, the total
energy-momentum tensor of GRMHD may be written as
\begin{align}
T^{ab}=\rho_0 h^* u^a u^b + p^* g^{ab}-b^a b^b\,,
\label{equation:EnergyMomentumTensorMHD}
\end{align}
with~$h^*=h+b^2/ \rho_0$,~$p^*=p+b^2/2$, and shorthand~$b^2=b^ab_a$.

The covariant system of evolution equations is given by the
conservation of the number of particles and the conservation of
energy momentum,
\begin{align}
\nabla_a(\rho_0 u^a)=0\,,\label{equation:ConsEquations}\\
\qquad \nabla_bT^{ab}=0\,,
\end{align}
plus the relevant Maxwell equations
\begin{align}
\nabla_b ({^*F}^{ab}-g^{ab}\phi )=- \frac{1}{\tau} n^a
\phi\,\label{equation:MaxwellEquations}, 
\end{align}
which are already augmented by the terms with the scalar field~$\phi$
to drive the Gauss constraint. Since~$b^a$ has only three free
components this equation now gives an evolution equation for~$b^a$
and~$\phi$. Elsewhere the notation~$\kappa=\tau^{-1}$ is employed. The
constant~$\tau$ is the timescale for the exponential driving toward
the Gauss constraint for the magnetic field. Typically~$\phi$ is set
to 0 in the initial and boundary conditions~\cite{DedKemKro02}.

\section{Hyperbolicity analysis of GRMHD with HDC}
\label{section:HypGRMHDDivCl}

Projecting
Eqs.~\eqref{equation:ConsEquations}-\eqref{equation:MaxwellEquations}
along the four velocity of the fluid~$u^a$ and perpendicular to it
by~$\gamu^a{}_b$, the nine evolution equations which determine the
time evolution of the GRMHD system with HDC are
\begin{align}
&\nabla_a(\rho_0 u^a)=0\,,\quad
\gamu_{ab}\nabla_cT^{bc}=0\,,\non \\ 
&\gamu_{ab}\nabla_c ({^*F}^{bc}-g^{bc}\phi)=-\frac{W}{\tau}V_a\phi\,,\non \\
&u_b\nabla_cT^{bc}=0\,,\quad  \ \
u_b \nabla_c ({^*F}^{bc}-g^{bc}\phi)=\frac{W}{\tau}\phi\,,\quad 
\label{equation:PDEMHD}
\end{align} 
supplemented with an EOS~\eqref{equation:EOS}. In the limit of~$\phi
\rightarrow 0$ we find the uppercase Gauss
constraint:~$\gamu^{bc}\nabla_bb_c=u_c\nabla_b {^*F}^{bc}=0$.

Taking Eq.~\eqref{equation:PDEMHD} and performing algebraic
manipulations similar to the investigation of other formulations of
GRMHD in Ref.~\cite{SchHilBug17}, we derive the evolution equations
for the pressure,
\begin{align}
\nabla_u p=& -\cs^2 \rho_0 h \gamu^{p}{}_c\gBi^{c e} \nabla_p \vhat_e 
+\frac{\kappa}{\rho_0}b^p\nabla_p\phi+S^{(p)}\,,
\label{equation:systemUpperCaseMHDp}
\end{align}
the boost vector,
\begin{align}
\gamu_{ab}&\gBi^{b c} \nabla_u \vhat_c =-\left(
\frac{b^pb_a}{\rho_0^2 h h^*}+\frac{\gamu^{p}{}_a}{\rho_0 h^*}\right)
\nabla_p p\non \\ 
&+ \left(\frac{2}{\rho_0 h^*}\gamu^{[b}{}_{a}b^{p]}
\gamu_{bc}+\frac{b_a}{\rho_0 h}\gamu^{p}{}_c \right)
\gBi^{c e}\nabla_{p}\pb_{e} \non \\ 
&+S^{(\mathbf{\vhat})}_a\,,
\label{equation:systemUpperCaseMHDvhat}
\end{align}
the magnetic field,
\begin{align}
\gamu_{ab}\gBi^{b c}\nabla_u \pb_c =&2\gamu_{ab}\gamu^{
[b}{}_c b^{p]}\gBi^{c e}\nabla_{p}\vhat_{e}\non \\ 
& -\gamu^p{}_a\nabla_p \phi +S^{(\mathbf{\pb})}_a
\,,
\label{equation:systemUpperCaseMHDperpb}
\end{align}
the specific internal energy,
\begin{align}
\nabla_u \varepsilon=& -\frac{ p}{\rho_0} \gamu^{p}{}_c\gBi^{c e} 
\nabla_p
\vhat_e+\frac{b^p}{\rho_0}\nabla_p \phi +S^{(\varepsilon)}\,
,\label{equation:systemUpperCaseMHDeps}
\end{align}
and finally the scalar field variable,
\begin{align}
\nabla_u \phi=& - \gamu^{p}{}_c\gBi^{c e} \nabla_p \pb_e
+S^{(\phi)}\,
.\label{equation:systemUpperCaseMHDphi}
\end{align}
The sources are given by 
\begin{align*}
&S^{(p)} =-\cs^2 W \rho_0 h \gamu^{d}{}_c\gBi^{c e} \nabla_d n_e
-\frac{\kappa W }{\tau \rho_0}(b^aV_a)\phi\,,\non \\
&S^{(\mathbf{\vhat})}_a= -W\gamu_{a b}\gBi^{b e}
\nabla_u n_e+ \frac{2 W}{\rho_0 h^*} \gamu^{[b}{}_{a}b^{e]}
V_bb^d\nabla_d n_e\\
&\quad \quad \quad +\frac{1}{\rho_0 h}b_a\left(WV^db^e-W(b^cV_c)
\gamu^{de}\right)\nabla_dn_e\,, \non \\
&S^{(\mathbf{\pb})}_a=2W \gamu_{ab}\gamu^{
  [b}{}_c b^{d]}\gBi^{c e}\nabla_d n_e \non\\\
  &\quad \quad \quad + 2W \gamu^{e}_{\ [a} V_{b]}b^b
\nabla_u n_e+\frac{W}{\tau}V_a\phi\,, \non \\
&S^{(\varepsilon)}=-\frac{W p}{\rho_0} \gamu^{d}{}_c\gBi^{c e}
\nabla_d n_e-\frac{W}{\tau \rho_0}(b^aV_a)\phi\,,\non\\
&S^{(\phi)}=-\left(WV^db^e-W(b^cV_c)\gamu^{de}\right)
\nabla_dn_e-\frac{W \phi}{\tau}.\non
\end{align*}
The auxiliary magnetic vector~$\pb_c$ is defined by the relation
\begin{align}
\gamu_{a c}\gBi^{c d}\nabla_{b}\pb_{d}:= &\
\gamu_{a c}\gBi^{c d}\nabla_{b}\hat{b}_{d}\non\\
&+V_a b_d \gBi^{de}\nabla_{b}\vhat_{e}\,.
\label{equation:DefinitionPerpb}
\end{align}
As usual, square brackets around indices denote antisymmetrization, so
that~$2\vhat^{[a}b^{b]}=\vhat^{a}b^{b}-\vhat^{b}b^{a}$. We have shown
explicitly that the set of
equations~\eqref{equation:systemUpperCaseMHDp}-\eqref{equation:systemUpperCaseMHDphi}
is, up to nonprincipal terms, which we have not carefully checked,
simply a linear combination of the formulation of GRMHD with HDC used
in numerical applications, see, for example,
Ref.~\cite{MoeMunFab14}. This verification can be found in the
notebook that accompanies the paper~\cite{HilSch18_WebTar}.

Writing
Eqs.~\eqref{equation:systemUpperCaseMHDp}-\eqref{equation:systemUpperCaseMHDphi}
in a vectorial form with state
vector~$\mathbf{U}=(p,\vhat_a,\pb_a,\varepsilon,\phi)^T$, we obtain,
in the notation of Ref.~\cite{SchHilBug17}, the principal part in the form
\begin{align}
\mathbf{B}^{\textrm u} \nabla_u \mathbf{U}=\mathbf{B}^p \nabla_p \mathbf{U} +
\boldsymbol{\mathcal{S}}\,.\label{equation:systemUpperCaseMHDMatrix}
\end{align}
Let~$S_a$ be an arbitrary unit spatial uppercase~$1$-form,~$S_aS^a=1$,
and~${\perpQ^{ b}{}\!_{a}}:=\gamu^b{}_a- S^bS_a$ be the associated
orthogonal projector. Let~$\s_a$ and~${\perpq^b{}\!_{a}}$ be their
lowercase projected versions,~$\s_a=\gamma^b{}_a
S_b$,~${\perpq^b{}\!_{a}}:=\gamma^b{}_a-\gBi^{bc}\s_c\s_a$.
Decomposing $\gamu^{b}{}_{a}$ and $\gamma^{b}{}_{a}$ against~$S_a$
and~$\s_a$, respectively,
Eq.~\eqref{equation:systemUpperCaseMHDMatrix} can be written as
\begin{align}
\left(\nabla_u\mathbf{U}\right)_{\shat,\,\ortAhat}\simeq\mathbf{P}^{S}
\left(\nabla_S\mathbf{U}\right)_{\shat,\,\ortBhat},
\label{equation:systemUpperCaseMHDMatrixPS}
\end{align}
where~$\simeq$ denotes equality up to nonprincipal terms and
uppercase spatial derivatives transverse to~$S^a$. The uppercase
principal symbol is~$\mathbf{P}^S=\mathbf{B}^S=$
\begin{align}
\begin{pmatrix}
0&-\cs^2 \rho_0 h & 0^{\ortb} & 0&  0^{\ortb}& 0& 
\frac{\kappa b^S}{\rho_0}\\
-\frac{\left(b^S\right)^2+\rho_0 h }{ \rho_0^2 hh^*}&0&0^{\ortb} &
 \frac{b^S}{\rho_0 h}
& -\frac{b^{\ortb}}{\rho_0 h^*}& 0& 0\\
-\frac{b^Sb_{\orta} }{ \rho_0^2 hh^*}& 0_{\orta}&0^{\ortb}{}_{\orta}&
\frac{b_{\orta}}{\rho_0 h} &
\frac{b^{S}}{\rho_0 h^*}\perpQ^{\ortb}\!\!_{\orta}&
0_{\orta}& 0_{\orta}\\
0& 0&0^{\ortb}& 0&0^{\ortb}& 0 & -1\\
0_{\orta}& -b_{\orta}&b^{S}\perpQ^{\ortb}\!\!_{\orta} & 0_{\orta}
& 0^{\ortb}{}_{\orta} & 0_{\orta} & 0_{\orta}\\
0& -\frac{p}{\rho_0}& 0^{\ortb} & 0& 0^{\ortb} & 0&
 \frac{b^S}{\rho_0}\\
0&0 &0^{\ortb} &-1 &0^{\ortb} &0 &0 \\ 
\end{pmatrix}\,\label{equation:PrincipalSymbolUpperCaseMHD}
\end{align}
with the state vector ordered as,
\begin{align}
(\delta \mathbf{U})_{\shat,\,\ortAhat}=(\delta p,
(\delta\vhat)_{\shat}, (\delta\vhat)_{\ortAhat}, 
(\delta \pb)_{\shat},
(\delta\pb)_{\ortAhat}, \delta\varepsilon, \delta\phi)^T.
\end{align}

The characteristic polynomial~$P_{\lambda}$ for the principal
symbol~\eqref{equation:PrincipalSymbolUpperCaseMHD} is calculated to
\begin{align}
&P_{\lambda}=\frac{\lambda}{ (\rho_0h^*)^2}(1-\lambda^2) 
P_{\text{Alfv\'en}} P_{\text{mgs}}\,,
\label{equation:CharPolUpperCaseMHD}
\end{align}
with the quadratic polynomial for Alfv\'en waves
\begin{align}
P_{\text{Alfv\'en}}=-\left(b^S\right)^2+\lambda^2\rho_0h^*
\label{equation:CharPolAlfvenUpperCaseMHD}
\end{align}
and the quartic polynomial for the magnetosonic waves
\begin{align}
P_{\text{mgs}}=\left(\lambda^2-1\right)
\left(\lambda^2b^2-\left(b^S\right)^2\cs^2\right)
+\lambda^2\left(\lambda^2-\cs^2
\right)\rho_0h\,.\label{equation:CharPolmgsUpperCaseMHD}
\end{align}
Comparing Eq.~\eqref{equation:CharPolmgsUpperCaseMHD} with our earlier
results for the flux-balance law formulation of
GRMHD in Ref.~\cite{SchHilBug17}, we see that the linear polynomial associated
with the Gauss constraint is replaced by the quadratic
polynomial~$1-\lambda^2$. The entropy, Alfv\'en, and slow and fast
magnetosonic uppercase eigenvalues remain the same, as before, and are
given by
\begin{align}
\lambda_{(\text e)}=&0\,,\quad \lambda_{(\text{a}\pm)}
=\pm\frac{b^S}{\sqrt{\rho_0h^*}}\,,\non \\ 
\lambda_{(\text{s}\pm)}&=\pm\sqrt{\zeta_{\text{S}}
-\sqrt{\zeta_{\text{S}}^2
-\xi_{\text{S}}}}\,,\non \\ \lambda_{(\text{f}\pm)}&
=\pm\sqrt{\zeta_{\text{S}}
+\sqrt{\zeta_{\text{S}}^2-\xi_{\text{S}}}}\,,
\end{align}
respectively, with shorthands
\begin{align}
\zeta_{\text{S}}=\frac{\left(
b^2+\cs^2\left[\left(b^S\right)^2+\rho_0h\right] \right)}{2\rho_0
h^*},\ \ \xi_{\text{S}}=\frac{\left(b^S\right)^2\cs^2 }{\rho_0 h^*}\,.
\end{align}
The remaining two speeds can be associated with the scalar field and
the longitudinal magnetic field~\cite{MoeMunFab14}, and are given by
\begin{align}
\lambda_{\pm}=\pm 1.
\end{align}
Since all uppercase eigenvalues have absolute value smaller than or
equal to 1, the relation~$|\lambda_{\text u}||V|<1$ is satisfied, so
we may analyze hyperbolicity independently of the
frame~\cite{SchHilBug17}. Therefore, we analyze the characteristic
structure of the principal symbol in the uppercase frame and the
result of the analysis applies directly to the numerically used system
(in the lowercase).

Continuing the characteristic analysis, we find the left entropy,
scalar field, and longitudinal magnetic field, Alfv\'en, and
magnetosonic eigenvectors being 

\begin{align}
&\begin{pmatrix}
-\frac{p}{\cs^2 \rho_0^2 h} & 0 & 0^{\orta} & \left(\rho_0
-\frac{\kappa p}{\cs^2 \rho_0 h}\right) \frac{b^S}{\rho_0^2} 
& 0^{\orta} & 1 & 0
\end{pmatrix}, \non \\
&\begin{pmatrix}
0 & 0 & 0^{\orta} & \pm 1 & 0^{\orta} & 0 & 1
\end{pmatrix},\non \\
&\begin{pmatrix}
0 & 0 & \mp\epsuS^{\orta \ortc} b_{\ortc}\sqrt{\rho_0 h^*}  & 
0 & -\epsuS^{\orta \ortc} b_{\ortc} & 0 & 0
\end{pmatrix},\non \\
&\begin{pmatrix}
\frac{\rho_0 h^*\left(\lambda_{(\text{m}\pm)}\right)^2 -b^2 }{\cs^2 \rho_0 h} &
\frac{\left(b^S\right)^2-\rho_0 h^*
\left(\lambda_{(\text{m}\pm)}\right)^2}{\lambda_{(\text{m}\pm)}}
&\frac{b^S b^{\orta}}{\lambda_{(\text{m}\pm)}}  & \mathcal{K} 
& b^{\orta} & \mathcal{L}
\end{pmatrix},\label{equation:LeftEigenvectorsMHDUpperCaseS}
\end{align}
respectively, where we defined the antisymmetric uppercase two- and
three-Levi-Civit\`a tensors as~${\epsuS^{\orta \ortb}=S_d\epsu^{d
\orta \ortb}=u_cS_d\,\perpQ^{\orta}\!_a\perpQ^{\ortb}\!_b
\epsilon^{cdab}}$. We employ furthermore the shorthands
\begin{align}
\mathcal{K}=&(b_{\perp}^2\cs^2+ 
\rho_0 h^*(\lambda_{(\text{m}\pm)})^2 -b^2)
\frac{(\kappa +\cs^2 \rho_0)b^S}{\cs^2\rho_0^2h(1-\cs^2)}\,;\non \\ 
\mathcal{L}=&\frac{\left(\rho_0 h^*\left(\lambda_{(\text{m}\pm)}\right)^2
-\left(b^S\right)^2\right)}{\left(\lambda_{(\text{m}\pm)}\right)^2}
\frac{\left(\kappa\left(\lambda_{(\text{m}\pm)}\right)^2+\cs^2\rho_0\right)
b^S}{(1-\cs^2)\rho_0^2 h \lambda_{(\text{m}\pm)}}\,.
\end{align}

The right eigenvectors can be computed and are presented in the same
order,
\begin{align}
&\begin{pmatrix}
0\\
0\\
0_{\ortb}\\
0\\
0_{\ortb}\\
1 \\
0
\end{pmatrix},\
\begin{pmatrix}
\mp \rho_0 h (\kappa  +\cs^2\rho_0)b^S\\
(\kappa + \rho_0) b^S\\
(1-\cs^2)\rho_0 b_{\ortb}\\
\pm (1-\cs^2)\rho_0^2 h\\
\mp (\kappa+\cs^2 \rho_0)b^S b_{\ortb}\\
\mp (\frac{\kappa p}{\rho_0} +p +(1-\cs^2)\rho_0 h)b^S \\
-(1-\cs^2)\rho_0^2 h
\end{pmatrix},\
\begin{pmatrix}
0\\
0\\
\mp  \frac{\epsuS_{\ortb \ortc}}{\sqrt{\rho_0 h^*}} b^{\ortc}\\
0\\
-\epsuS_{\ortb \ortc} b^{\ortc}\\
0\\
0
\end{pmatrix},\non\\
&\begin{pmatrix}
\frac{\cs^2 \rho_0^2 h}{p}\\ - \frac{\rho_0
  \lambda_{(\text{m}\pm)}}{p}\\ \frac{\rho_0 \lambda_{(\text{m}\pm)}}{p
  b^{S} b_{\perp}^2}\left[\left(b^S\right)^2+\rho_0
  h^*\left((\lambda_{(\text{m}\pm)})^2-2\zeta_{\text S}\right) \right]
b_{\ortb}\\ 0\\ \frac{\rho_0}{ b_{\perp}^2p} \left[b^2+\rho_0
  h^*\left((\lambda_{(\text{m}\pm)})^2-2\zeta_{\text S}\right) \right]b_{\ortb}\\
1\\
0
\end{pmatrix}.\label{equation:RightEigenvectorsMHDUpperCaseS}
\end{align}

We have introduced in the magnetosonic eigenvectors the orthogonal
magnetic field vector as~$b_{\perp}^a=\perpQ_{\ b}^{a} b^b$
with~$b_{\perp}^2=b_{\perp}^a b^{\perp}_a=b^\orta b_\orta$. As for the
 {\it prototype algebraic constraint free formulation} treated
in Ref.~\cite{SchHilBug17}, rescaled versions of the left
eigenvectors~\eqref{equation:LeftEigenvectorsMHDUpperCaseS} and right
eigenvectors~\eqref{equation:RightEigenvectorsMHDUpperCaseS} can be
derived. They form complete sets of nine linearly independent
eigenvectors under type I, type II, and type II$'$
degeneracies~\cite{AntMirMar10,Sch18}. The rescaling can be found in the
notebook provided in Ref.~\cite{HilSch18_WebTar}. Thus, as long
as~$p=p(\rho_0,\varepsilon)>0$ and~$0<\cs<1$ hold, the formulation of
GRMHD with HDC as given above forms a strongly
hyperbolic system of equations.

In the limit of~$\cs\rightarrow 1$, it can be shown, that the fast
magnetosonic waves collide pairwise with the waves associated to the
scalar field and longitudinal magnetic field, in the case of which the
system is only weakly hyperbolic. The limiting procedure can be found
in the provided notebook. This is a consequence of taking the
divergence cleaning to happen at the speed of light. By the simple
replacement~$\phi \rightarrow c_{\phi}^{-2} \phi$, $c_{\phi}>0$ in
Eq.~\eqref{equation:systemUpperCaseMHDphi}, the divergence cleaning
speed becomes~$\lambda_{\pm}= \pm c_{\phi}$. For~$c_{\phi}>1$, strong
hyperbolicity is also guaranteed in the limiting case~$\cs=1$. This
strategy does however place a nontrivial upper limit on the speed of
flows that can be managed with the method, as strong hyperbolicity
will break down for sufficiently fast flows. See
Ref.~\cite{SchHilBug17} for details. By modifying the lowercase
equations directly it may be possible to avoid this shortcoming, too.

Finally, we want to present the uppercase rescaled characteristic
variables for GRMHD with HDC. They are valid for all
degeneracies, and are given by
\begin{align}
\hat{\text{U}}_{\text{e}}=&\delta \varepsilon-\frac{p}{\cs^2 \rho_0^2 h}\delta
p+\left(\rho_0
-\frac{\kappa p}{\cs^2 \rho_0 h}\right)
\frac{b^S}{\rho_0^2}(\delta \pb)_{{\shat}} \,,\non \\ 
\hat{\text{U}}_{\pm}=&\delta \phi \pm (\delta
\pb)_{\hat{\s}}\,,\non\\ 
\hat{\text{U}}_{\text a \pm}=&\pm\epsuS^{\orta
\ortc} \sqrt{\rho_0 h^*}\frac{b^{\perp}_{\ortc}}{|b_{\perp}|}
(\delta \vhat)_{{\ortAhat}}+\epsuS^{\orta \ortc}
\frac{b^{\perp}_{\ortc}}{|b_{\perp}|} (\delta \pb)_{{\ortAhat}}\,, \non \\ 
\hat{\text{U}}_{\text m_1 \pm}=&\frac{\mathcal{H}(\lambda^2-1)}{\rho_0 h}\delta p+
(1-\cs^2)\mathcal{H}\lambda(\delta\vhat)_{\hat{\s}}\non
\\ &+ \left(\frac{b^S}{\lambda}\right)\frac{b_{\perp}^{\orta}}{|b_{\perp}|}
(\delta \vhat)_{\ortAhat} -\frac{\mathcal{H}(\kappa +\cs^2 \rho_0)b^S}
{\rho_0^2 h}(\delta \pb)_{{\shat}} \non \\
&+ \frac{b_{\perp}^{\orta}}{|b_{\perp}|}
(\delta \pb)_{\ortAhat} + \left(\frac{b^S
}{\lambda}\right)\frac{\mathcal{H}(\kappa\lambda^2 +\cs^2 \rho_0)}
{\rho_0^2 h}\delta \phi\,,\non\\ 
\hat{\text{U}}_{\text m_2
  \pm}=& \frac{1}{\cs^2\rho_0 h}\delta p+
\frac{(1-\cs^2)\lambda}{\cs^2(\lambda^2-1)}(\delta\vhat)_{\hat{\s}}
+\left(\frac{b^S}{\lambda}\right)\mathcal{F}^{\orta}(\delta
\vhat)_{\ortAhat} \non\\ 
&+\left(\frac{b^S}{\lambda}\right) \frac{\lambda(\kappa + \cs^2 \rho_0)}
{\cs^2(1-\lambda^2)\rho_0^2 h}(\delta \pb)_{{\shat}}
+ \mathcal{F}^{\orta}(\delta \pb)_{\ortAhat} \non \\
&-\left(\frac{b^S}{\lambda}\right)
\frac{(\kappa \lambda^2 + \cs^2 \rho_0)}
{\cs^2(1-\lambda^2)\rho_0^2 h}\delta \phi \,,
\label{equation:CharVarUpperGRMHDDC}
\end{align}
with~$\{\text{m}_1,\text{m}_2\}$ equal to $\{\text{s},\text{f}\}$ or
$\{\text{f},\text{s}\}$. The abbreviations in
Eq.~\eqref{equation:CharVarUpperGRMHDDC} are given by
\begin{align}
\mathcal{H}=&\frac{|b_{\perp}|}{\cs^2 -\lambda^2 }\,,\\
\mathcal{F}^{\orta}=&
\frac{b_{\perp}^{\orta}}{(\rho_0 h^*\lambda^2-b^2)}\,,
\end{align}
where for type II and even for type II$'$ degeneracy we take~$\Qa^a$
and~$\Qb^a$ such that in the degenerate limit we have
\begin{align}
\frac{b^{\perp}_{\ortc}}{|b_{\perp}|}  =
&\frac{1}{\sqrt{2}}(\Qa_{\ortc}+\Qb_{\ortc})\,,
\label{equation:degenerateperpb} \\
\mathcal{H}=&0\,,\\
\mathcal{F}^{\orta}=&0^{\orta}\,.\label{equation:degenerateFa}
\end{align}
For further explanations concerning degeneracies and rescaling, see
also Ref.~\cite{Sch18}.

Using the recovery procedure given in Ref.~\cite{SchHilBug17}, the
lowercase characteristic quantities such as eigenvalues and
eigenvectors can be derived. The calculation can be found in the
notebook~\cite{HilSch18_WebTar}, but results in rather long
expressions which we suppress here. Both the lowercase left
magnetosonic eigenvectors and the lowercase right eigenvectors
associated with the scalar field and longitudinal magnetic field
eigenvalues have a particularly complicated structure, for which a
useful simplification seems difficult. In applications it may
therefore be appropriate to compute the characteristics numerically.

\section{Discussion of formulations of GRMHD with VP}

The formulations of GRMHD we have thus far considered use the magnetic
field as an evolved variable. Another possibility is to introduce the
four-vector potential
instead~\cite{GiaRezBai10,EtiPasLiu12,EtiPasHaa15}.  In practice, the
potential is then~$3+1$ decomposed. Such formulations have the
advantage that the Gauss constraint is satisfied by construction, and
in this sense can be considered a type of constrained- rather than
free-evolution. On the other hand one obtains a system of equations
which is {\it a priori} not, from the PDE point of view, minimally
coupled to the gravitational field equations. The resulting evolution
equations for the GRMHD variables are moreover themselves not in
first-order form, but rather first order in time and second order in
space, {\it and} there is an additional gauge degree of
freedom. Different choices in this freedom may have different PDE
properties as the principal part of the evolution system is
altered. We follow Ref.~\cite{EtiPasLiu12} and focus on the Lorenz
gauge, but similar comments hold elsewhere. Strong hyperbolicity
of first order in time, second order in space systems can be
defined~\cite{GunGar05,HilRic13a} by the requirement that there exists
a first-order reduction which satisfies the definition given for first
order PDEs in Sec.~\ref{section:MathBackground}. Therefore, we must
reduce the governing system of equations as in
Eq.~\eqref{equation:PDEsystem}, by introducing reduction
variables. There are two natural ways to go about this.

The first, naive, possibility is to introduce reduction
constraints~$c_{ab}=d_{ab}-\gamma^c{}_a\gamma^d{}_b\p_cA_d$, which
should vanish, for the lowercase spatial derivatives of the lowercase
spatial part of the vector potential~$A_a$, and likewise for the
electric potential. The reduction variables~$d_{ab}$ should satisfy
also the ordering constraint,
\begin{align}
c_{abc}=\gamma^d{}_a\gamma^e{}_b\gamma^f{}_c\p_{[d}c_{e]f}
=\gamma^d{}_a\gamma^e{}_b\gamma^f{}_c\p_{[d}d_{e]f}=0\,,
\end{align}
and similarly for the electric potential reduction variables. The
reduction constraints must then be added to the equations of motion to
remove all second spatial derivatives. Besides that, both the
reduction and ordering constraints can be added freely to try and find
a hyperbolic reduction. Such a reduction does not use the special
structure of the Maxwell equations, does not utilize the fact that the
original system satisfies the Gauss constraint by construction, and is
not minimally coupled to the evolution equations for the geometric
variables. Worse, the resulting principal symbol does not have a clear
structure, which makes the analysis very difficult.

The less obvious option is to bring back the magnetic field as a
reduction variable for the curl of the spatial vector potential by
defining a reduction constraint,
\begin{align}
C_a=\epsilon_a{}^{bc}D_b A_c - B_a\,.
\end{align}
In this reduction we need not introduce a reduction variable to the
electric potential as it appears with at most one spatial
derivative. Part of the analog of the ordering constraint in such a
reduction turns out to be simply the Gauss constraint,
\begin{align}
C=-D_aC^a=D_aB^a\,.
\end{align}
A generic PDE system does not allow a reduction of this type, in which
new variables that only capture {\it part} of the spatial derivatives
are introduced. Due to the gauge freedom of the Maxwell equations
however the `longitudinal' part of the vector potential does not
appear elsewhere in the remaining equations of motion, and so we can
close the evolution system using only~$B_a$. Note that such a
restricted reduction does have consequences on the norms in which
rigorous estimates would be demonstrated, and also that as usual first
derivatives of the metric here are nonprincipal.

Ultimately we end up with evolution equations for the matter variables
which are minimally coupled to the Einstein equations. Naively writing
out the lowercase principal symbol of the matter variables we can
obtain moreover a block-diagonal structure,
\begin{align}
\mathbf{P}^s=\begin{pmatrix}
\mathbf{A} & \mathbf{0} \\
0 & \mathbf{B} \\
\end{pmatrix},
\end{align}
where block~$\mathbf{A}$ denotes the principal symbol of the system of
evolution equations of the spatial part of the vector potential and
the electric potential, whereas~$\mathbf{B}$ can be rendered identical
to the principal symbol of the {\it prototype algebraic constraint
  free formulation} of GRMHD investigated in Ref.~\cite{SchHilBug17}. Here,
crucially, we rely on the fact that, as it is not to be used in
applications, this formal first-order reduction need not be of a
flux-balance form, and therefore we can add the ordering
constraint~$C$ as desired. The upper right block vanishes trivially
and the lower left block vanishes by appropriate choice of reduction.
We showed already that {\it prototype algebraic constraint free
  formulation} of GRMHD is strongly hyperbolic in the lowercase frame,
with an EOS of the form~\eqref{equation:EOS} and~$0<\cs\leq 1$, so all
that remains is to show that the block~$\mathbf{A}$ satisfies the
conditions for strong hyperbolicity. This was done already
in Ref.~\cite{EtiPasLiu12}, but with the use of the reduction
variable~$B_a$ we can give a slightly cleaner treatment. The lowercase
principal symbol can be read off from,
\begin{align}
\nabla_n\Phi&\simeq -\gamma^{pe}\nabla_p A_e   \,,\\
\gamma^b{}_a\nabla_nA_b&\simeq -\gamma^p{}_a\nabla_p\Phi
\,.\label{equation:VecPot}
\end{align}
Note that in Eq.~\eqref{equation:VecPot} the term~$D_aA_b-D_bA_a$ is
written in terms of the reduction variable~$B_a$ and does not
contribute to the principal part. Let~$s^a$,~$s_as^a=1$, be a unit
spatial lowercase vector and~$q^a{}_b=\gamma^a{}_b-s^as_b$ be the
orthogonal projector. The characteristic variables associated with
this block are hence
\begin{align}
\delta\Phi \mp(\delta A)_s\,,
\end{align}
with speeds~$\pm 1$, respectively, and
\begin{align}
 (\delta A)_A\,,
\end{align}
with speed~$0$ for the two orthogonal directions to~$s^a$. The
calculation is provided in a notebook that accompanies the
paper~\cite{HilSch18_WebTar}.

\section{Conclusion}\label{section:Conclusion}

In previous work~\cite{SchHilBug17} we examined two formulations of
ideal GRMHD, and showed that a formulation similar to that studied in
Refs.~\cite{AniPen87,Ani90a}, which we call the {\it prototype
  algebraic constraint free formulation} is strongly
hyperbolic. Unfortunately, this formulation is not in the flux-balance
law form desirable for the application of standard numerical
methods. Turning to GRMHD in flux-conservative form, we found the
system to be only weakly hyperbolic. This formulation of GRMHD hence
has an ill-posed initial value problem. Fortunately, two popular,
applicable, alternative formulations of GRMHD were left untreated by
that analysis. Presently, we have addressed this shortcoming with the
outcome first, that formulations of GRMHD with
HDC~\cite{DedKemKro02,AndHirLie06,MoeMunFab14} are indeed strongly
hyperbolic as long as the sound speed is suitably
bounded~$0<\cs<1$. In fact, it is straightforward to achieve
hyperbolicity also in the case~$c_s=1$ by changing the speed of the
cleaning in the formulation. Second, we have shown that by a careful
reduction to first-order, formulations of GRMHD with
VP~\cite{EtiPasLiu12} can also be rendered strongly hyperbolic
whenever~$0<\cs\leq1$. The latter result is a corollary of strong
hyperbolicity of the prototype algebraic constraint free
formulation. Here we have discussed only the Lorenz gauge choice, but
our results carry over trivially to generalized Lorenz gauge, in which
there is a modification by source terms, and a natural treatment will
be very similar in other cases, too.

Both HDC and the VP formulations were introduced as strategies to
control Gauss-constraint violation in applications. Another popular
approach, called constrained transport
(CT)~\cite{EvaHaw88,BalSpi99,Tot00}, uses a carefully constructed
discretization so that in a particular approximation the constraint is
identically satisfied. There is some subtlety in precisely what
continuum PDE should be analyzed given such a constrained evolution,
but supposing that the constraints are identically satisfied, they may
again be added arbitrarily to the evolution equations, and
strong hyperbolicity can again be achieved, in the restricted,
constraint-satisfying phase space, as a corollary of hyperbolicity of
the prototype algebraic constraint free formulation.

In Ref.~\cite{SchHilBug17} we discussed two minimally coupled
formulations of {\it resistive} GRMHD with HDC, one with and one
without the evolution of the charge density~$q$. Both were found to be
only weakly hyperbolic. A natural question is therefore whether the
use of the VP approach could cure this problem. Replacing the
divergence cleaning variables by~$A_a$ and~$\Phi$, and making a
minimally coupled first-order reduction as we did for GRMHD, one
arrives with a lower block triangular structure in the principal
symbol, with the lower-right block~$\mathbf{C}$ being precisely a
sub-block of the principal symbol of the original formulation of
RGRMHD. Neither of the original two formulations were strongly
hyperbolic because~$\mathbf{C}$ was not diagonalizable. Consequently,
the vector potential formulations are also not strongly
hyperbolic. Thus, at least if we insist on taking only minimally
coupled first-order reductions, use of a VP reformulation of RGRMHD
does nothing to circumvent weak hyperbolicity of RGRMHD.

For numerical applications we therefore have the clear conclusion
that, by the fundamental requirement of well-posedness, HDC and VP
formulations (and likely also CT schemes) are preferred over their
older variant which should henceforth be avoided. From the PDE point
of view it is, at this stage, difficult to choose between the favored
formulations. One might be tempted to argue in favor of the vector
potential formulation, as indeed it is true that there the
characteristic structure, inherited from the prototype algebraic
constraint free formulation, is simpler, but this is not a principle
advantage. In the future it is hoped that the characteristic structure
uncovered by our analysis can be put to good use in numerical work in
both systems.

\acknowledgments

We are grateful to Sebastiano Bernuzzi and Bruno Giacomazzo for useful
discussions and comments. This work was partially supported by the FCT
(Portugal) IF Program No. IF/00577/2015 and the GWverse COST action
Grant No.~CA16104.


\begin{thebibliography}{10}

\bibitem{BauSha10x}
Thomas Baumgarte and Stuart Shapiro.
\newblock {\em Numerical Relativity}.
\newblock Cambridge University Press, Cambridge, England, 2010.

\bibitem{Shi16}
Masaru Shibata.
\newblock {\em {Numerical Relativity}}.
\newblock World Scientific, Singapore, 2016.

\bibitem{SchHilBug17}
Andreas Schoepe, David Hilditch, and Marcus Bugner.
\newblock {Revisiting Hyperbolicity of Relativistic Fluids}.
\newblock {\em Phys. Rev. D}, 97:123009, 2018.

\bibitem{HilRic13}
David Hilditch and Ronny Richter.
\newblock {Hyperbolicity of Physical Theories with Application to General
  Relativity}.
\newblock {\em Phys. Rev. D}, 94(4):044028, 2016.

\bibitem{Hil15}
David Hilditch.
\newblock {Dual Foliation Formulations of General Relativity}.
\newblock {arXiv:1508.02071}.

\bibitem{HilRui16}
David Hilditch and Milton Ruiz.
\newblock {The initial boundary value problem for free-evolution formulations
  of General Relativity}.
\newblock {\em Classical Quantum Gravity}, 35:015006, 2018.

\bibitem{HilHarBug16}
David Hilditch, Enno Harms, Marcus Bugner, Hannes R{\"u}ter, and Bernd
  Br{\"u}gmann.
\newblock {The evolution of hyperboloidal data with the dual foliation
  formalism: Mathematical analysis and wave equation tests}.
\newblock {\em Classical Quantum Gravity}, 35(5):055003, 2018.

\bibitem{AntZanMir05}
Luis Ant{\'o}n, Olindo Zanotti, Joan.~A. Miralles, Jos{\'e}~M. Mart{\'i},
  Jos{\'e}~M. Ib{\'a}{\~n}ez, Jos{\'e}~A. Font, and Jos{\'e}~A. Pons.
\newblock Numerical 3+1 general relativistic magnetohydrodynamics: a local
  characteristic approach.
\newblock {\em Astrophys. J.}, 637:296, 2006.

\bibitem{xAct_web_aastex}
Jos{\'e}~M. Mart{\'i}n-Garc{\'i}a.
\newblock x{A}ct: Tensor computer algebra,
\newblock \url{http://www.xact.es/}.

\bibitem{HilSch18_WebTar}
{\url{https://centra.tecnico.ulisboa.pt/~hilditch/Hydro_DF_V2.tgz}}.

\bibitem{GusKreOli95}
Bertil Gustafsson, Heinz-Otto Kreiss, and Joseph Oliger.
\newblock {\em Time Dependent Problems and Difference Methods}.
\newblock Wiley, New York, 1995.

\bibitem{SarTig12}
Olivier Sarbach and Manuel Tiglio.
\newblock Continuum and discrete initial-boundary value problems and einstein's
  field equations.
\newblock {\em Living Reviews in Relativity}, 15(9), 2012.

\bibitem{Hil13}
David Hilditch.
\newblock {An Introduction to Well-posedness and Free-evolution}.
\newblock {\em Int. J. Mod. Phys. A}, 28:1340015, 2013.

\bibitem{Ani90a}
A.~M. Anile.
\newblock {\em Cambridge Monographs on Mathematical Physics.} Cambridge University Press, Cambridge, England, 1990.

\bibitem{AntMirMar10}
Luis Ant{\'o}n, Juan~A. Miralles, Jos{\'e}~M. Mart{\'i}, Jos{\'e}~M.
  Ib{\'a}{\~n}ez, Miguel~A. Aloy, and Petar Mimica.
\newblock {Relativistic Magnetohydrodynamics: Renormalized eigenvectors and
  full wave decomposition Riemann solver}.
\newblock {\em Astrophys. J. Suppl.}, 188:1--31, 2010.

\bibitem{MoeMunFab14}
Philipp M{\"o}sta, Bruno~C. Mundim, Joshua~A. Faber, Roland Haas, Scott~C.
  Noble, Tanja Bode, Frank L{\"o}ffler, Christian~D. Ott, Christian Reisswig,
  and Erik Schnetter.
\newblock {GRHydro: A new open source general-relativistic magnetohydrodynamics
  code for the Einstein Toolkit}.
\newblock {\em Classical Quantum Gravity}, 31:015005, 2014.

\bibitem{AlcDegSal09}
Miguel Alcubierre, Juan~Carlos Degollado, and Marcelo Salgado.
\newblock {The Einstein-Maxwell system in 3+1 form and initial data for
  multiple charged black holes}.
\newblock {\em Phys.Rev. D}, 80:104022, 2009.

\bibitem{DedKemKro02}
A.~Dedner, F.~Kemm, D.~Kr{\"oner}, C.-D. Munz, T.~Schnitzer, and M.~Wesenberg.
\newblock {Hyperbolic Divergence Cleaning for the MHD Equations}.
\newblock {\em J. Comput. Phys.}, 175:645--673, 2002.

\bibitem{Sch18}
Andreas Schoepe.
\newblock {\em {On the Hyperbolicity of Evolution Equations for Relativistc
  Fluids}}.
\newblock PhD thesis, {University of Jena}, 2018.

\bibitem{GiaRezBai10}
Bruno Giacomazzo, Luciano Rezzolla, and Luca Baiotti.
\newblock {Accurate evolutions of inspiralling and magnetized neutron-stars:
  equal-mass binaries}.
\newblock {\em Phys. Rev. D}, 83:044014, 2011.

\bibitem{EtiPasLiu12}
Zachariah~B. Etienne, Vasileios Paschalidis, Yuk~Tung Liu, and Stuart~L.
  Shapiro.
\newblock {Relativistic MHD in dynamical spacetimes: Improved EM gauge
  condition for AMR grids}.
\newblock {\em Phys. Rev. D}, 85:024013, 2012.

\bibitem{EtiPasHaa15}
Zachariah~B. Etienne, Vasileios Paschalidis, Roland Haas, Philipp M{\"o}sta,
  and Stuart~L. Shapiro.
\newblock {IllinoisGRMHD: An Open-Source, User-Friendly GRMHD Code for
  Dynamical Spacetimes}.
\newblock {\em Classical Quantum Gravity}, 32:175009, 2015.

\bibitem{GunGar05}
Carsten Gundlach and Jose~M. Mart{\'i}n-Garc{\'i}a.
\newblock Hyperbolicity of second-order in space systems of evolution
  equations.
\newblock {\em Classical Quantum Gravity}, 23:S387--S404, 2006.

\bibitem{HilRic13a}
David Hilditch and Ronny Richter.
\newblock {Hyperbolicity of High Order Systems of Evolution Equations}.
\newblock {\em J. Hyperbolic Differ. Equations}, 12(1), 2015.

\bibitem{AniPen87}
A.~M. Anile and S.~Pennisi.
\newblock {\em On the mathematical structure of test relativistic
  magnetofluiddynamics.}
\newblock  Ann. l'I.H.P. Phys. th{\'e}orique, 46(1):27--44, 1987.

\bibitem{AndHirLie06}
Matthew Anderson, Eric Hirschmann, Steven~L. Liebling, and David Neilsen.
\newblock Relativistic {MHD} with {Adaptive} {Mesh} {Refinement}.
\newblock {\em Classical Quantum Gravity}, 23:6503--6524, 2006.

\bibitem{EvaHaw88}
Charles~R. Evans and John~F. Hawley.
\newblock Simulation of magnetohydrodynamic flows - a constrained transport
  method.
\newblock {\em Astrophys. J.}, 332:659--677, 1988.

\bibitem{BalSpi99}
Dinshaw~S. Balsara and Daniel~S. Spicer.
\newblock A staggered mesh algorithm using high order godunov fluxes to ensure
  solenoidal magnetic fields in magnetohydrodynamic simulations.
\newblock {\em J. Comput. Phys.}, 149:270--292, 1999.

\bibitem{Tot00}
G.~Toth.
\newblock The div b=0 constraint in shock-capturing magnetohydrodynamics codes.
\newblock {\em J. Comput. Phys.}, 161:605--652, 2000.

\end{thebibliography}


\end{document}